\documentclass[10pt,prl,twocolumn,floatfix,tightenlines,showpacs,showkeys,preprintnumbers,altaffilletter, nofootinbib, superscriptaddress, longbibliography]{revtex4-1}
\usepackage{graphicx,mathtools,tabu}
\usepackage{epstopdf}
\usepackage{amsmath}
\usepackage{amsfonts}
\usepackage[colorlinks=true,citecolor=red,linkcolor=blue,breaklinks=true]{hyperref}
\usepackage[dvipsnames]{xcolor}
\usepackage{accents}
\usepackage[figure, table]{hypcap}
\usepackage{amssymb}
\usepackage{appendix}
\usepackage{comment}
\usepackage{bbold}
\usepackage{color}
\usepackage{slashed}
\usepackage{subfigure}
\usepackage{setspace}
\usepackage{footnote}
\usepackage{multirow}
\usepackage{braket}
\usepackage[normalem]{ulem}
\usepackage[utf8]{inputenc}
\usepackage{mathrsfs}
\usepackage{longtable}
\usepackage{enumitem}

\date{\today}

\begin{document}
\preprint{MPP-2020-79}
\title{
Fast flavor conversions in supernovae: the rise of mu-tau neutrinos
}

\author{Francesco Capozzi}
\email{capozzi@mppmu.mpg.de}
\affiliation{Max-Planck-Institut f\"{u}r Physik (Werner-Heisenberg-Institut), F\"{o}hringer Ring 6, 80805 M\"{u}nchen, Germany}

\author{Madhurima Chakraborty}
\email{madhu176121012@iitg.ac.in}
\affiliation{Indian Institute of Technology,
Guwahati, Assam-781039, India}

\author{Sovan Chakraborty}
\email{sovan@iitg.ac.in}
\affiliation{Indian Institute of Technology, 
Guwahati, Assam-781039, India}

\author{Manibrata Sen}
\email{manibrata@berkeley.edu}
\affiliation{Northwestern University, Department of Physics \& Astronomy, 2145 Sheridan Road, Evanston, IL 60208, USA}
\affiliation{Department of Physics, University of California Berkeley, Berkeley, California 94720, USA}

\begin{abstract}
Neutrinos in a core-collapse supernova can undergo fast flavor conversions with a possible impact on the explosion mechanism and nucleosynthesis. We perform the first non-linear simulations of fast conversions in the presence of three neutrino flavors. The recent supernova simulations with muon production call for such an analysis, as they relax the standard $\nu_{\mu,\tau}=\bar{\nu}_{\mu,\tau}$ (two-flavor) assumption. Our results show the significance of muon and tau lepton number angular distributions, together with the traditional electron lepton number ones. Indeed, our three-flavor results are potentially very different from two-flavor ones. 
These results strengthen the need to further investigate the occurrence of fast conversions in supernova simulation data, including the degeneracy breaking of mu and tau neutrinos.
\end{abstract}

\maketitle

\emph{Introduction.\,}-- Neutrinos streaming out of a supernova (SN) encounter a large density of ambient neutrinos and antineutrinos.
Forward-scattering off this surrounding neutrino bath has been shown to cause self-induced effects, leading to collective flavor oscillations, where neutrinos with different oscillation frequencies change their flavors in a coherent fashion
(see~\cite{Duan:2010bg,Mirizzi:2015eza,Chakraborty:2016yeg,Horiuchi:2017sku} for a comprehensive review). Recently, it was shown that under certain conditions, these collective oscillations can be ``fast", growing with the extremely large neutrino density $n_\nu$, while being independent of the neutrino mass-squared difference $\Delta m^2$ or the neutrino energy $E$~\cite{Sawyer:2015dsa, Chakraborty:2016lct,Dasgupta:2016dbv}. These fast flavor conversions (FFC) can occur close to the region of neutrino free-streaming, and can grow as large as $10^5$ times faster than the vacuum oscillations. 

A host of studies \cite{Sawyer:2005jk,Sawyer:2008zs,Sawyer:2015dsa,Chakraborty:2016lct,Dasgupta:2016dbv,Izaguirre:2016gsx,Capozzi:2017gqd, Dighe:2017sur, Dasgupta:2017oko,Dasgupta:2018ulw,Abbar:2018shq,Azari:2019jvr,Johns:2019izj,Glas:2019ijo,Shalgar:2019qwg,Bhattacharyya:2020dhu} in the past few years suggest that a necessary, though not sufficient, condition for the existence of these fast instabilities is the presence of a \emph{zero-crossing} in the angular distribution of the neutrino electron lepton number (ELN), i.e., the difference between the electron neutrino and the antineutrino angular emission spectra should go through a zero for some emission angle. Though neutrinos of other flavors are also present in the SN environment, ELN is considered to be the only driving quantity, as these studies are performed assuming an effective two-flavor scenario ($e$ and $x$ flavor, where $x=\mu,\,\tau$ or a linear combination of both). This is due to the fact that in the absence of muons and taus in traditional SN simulations, the heavy lepton neutrinos have identical microphysics and similar number density  
($n_{\nu_\mu}=n_{\nu_\tau}=n_{\bar{\nu}_\mu}=n_{\bar{\nu}_\tau}=n_{\nu_x}$) for all emission angles.
Crossings are naively expected to occur near the neutrino free-streaming zone and require a comprehensive study of the interplay of collisions and fast conversions, thereby necessitating the use of quantum kinetic equations~\cite{Vlasenko:2013fja,Volpe:2015rla,Capozzi:2018clo,Richers:2019grc}.
Recently, dedicated studies have performed a thorough scan of existing data from SN hydrodynamical simulations. In \cite{Abbar:2019zoq} it was shown that crossings in the ELN, sometimes related to the lepton number emission self-sustained asymmetry (LESA) \cite{Tamborra:2014aua,Glas:2018vcs}, are present both in the convective layer of the proto-neutron star (PNS) and outside the neutrino decoupling region. Such results have been confirmed with further insights from \cite{Glas:2019ijo}. Furthermore, in \cite{Morinaga:2020nmf}, it was claimed that coherent neutrino-nucleus scattering in the pre-shock region can lead to a tiny zero crossing in the backward direction, leading to FFC at a distance of $\mathcal{O}(100)\,{\rm km}$ from the PNS.

The presence of these recently discovered FFCs well inside a SN can lead to paradigm changes in our understanding of the explosion dynamics, requiring a more detailed analysis of the different approximations going into the studies. In particular, the above results are based on the effective two-flavor setup. This set up has severe limitations as the $\nu_x$ and $\bar\nu_x$ flux would differ naturally if nucleon recoil effects, implying different neutral-current scattering cross sections, are taken into account. Moreover, it would be also natural to expect that the high temperatures during the accretion phase would create muons in the nascent neutron stars. In fact, \cite{Bollig:2017lki} has shown that the addition of muons can enhance neutrino energy deposition to the stalled shockwave, leading to a successful explosion. Muon production in a SN \cite{Bollig:2017lki} can create differences in the heavy lepton flavor neutrino fluxes, thereby necessitating the inclusion of three-flavor effects. The oscillation treatment of such a scenario has been pointed out in \cite{Airen:2018nvp} and recently, \cite{Chakraborty:2019wxe} has done a detailed analysis of fast conversions to three neutrino flavors. Using a linear stability analysis, it demonstrated the possibility of altering the instability growth rates obtained in the standard two-flavor setup.

Going a step forward with respect to current literature, we, {\it{for the first time}}, perform a {\it{fully non-linear}} computation of FFCs in the presence of three neutrino flavors. Motivated by the difference in the heavy lepton flavor neutrino fluxes observed in \cite{Bollig:2017lki}, we propose simple toy models and demonstrate that it is not only the ELN, but also $\mu\,{\rm LN}$ and $\tau\,{\rm LN}$ that drives the onset of these rapid conversions. In fact, it is the difference of any of the two-flavor lepton numbers that goes into the evolution equations \cite{Chakraborty:2019wxe}. 
This emphasizes the {\it{incompleteness}} of a two-flavor analysis, 
especially while analyzing the presence of FFC in the pre-shock region, where tiny crossings in the ELN can get erased by the $\mu{\rm LN}$ and or $\tau{\rm LN}$. 
Our results suggest the importance of such a fully non-linear three-flavor study of FFC with detailed hydrodynamic SN simulations, including muon production, to gauge its impact on SN dynamics.
 
\emph{Equations of motion.\,}-- The dynamics of the neutrino occupation number matrices $\rho_{{\bf p}, {\bf x},t}$ for momentum ${\bf p}$ at position ${\bf x}$ and time $t$ is governed by the equations of motion (EoMs)~\cite{Sigl:1992fn}
\begin{equation}
i\left(\partial_t + {\bf v}_{\bf p} \cdot \nabla_{\bf x}\right) \rho_{{\bf p}, {\bf x},t} 
= [\Omega_{{\bf p}, {\bf x},t}, \rho_{{\bf p}, {\bf x},t}] 
\,\ ,
\label{eq:eom}
\end{equation}
where the left-hand-side accounts for time and spatial dependence of $\rho_{{\bf p}, {\bf x},t}$. The Hamiltonian $\Omega_{{\bf p}, {\bf x},t}$ on the right-hand-side consists of the vacuum term $(\Delta m^2/2\,E)$, the Mikheyev-Smirnov-Wolfenstein matter potential depending on the background charged lepton number density and the $\nu-\nu$ self-interaction potential, \,\mbox{$\int d^3{\bf q}/(2\pi)^3 (1-{\bf v}_{\bf p}\cdot {\bf v}_{\bf q})({\rho}_{{\bf q}, {\bf x},t}-{\bar\rho}_{{\bf q}, {\bf x},t})$}~\cite{Mirizzi:2015eza}

In the context of fast oscillations, the vacuum term does not play a major role, except for seeding the oscillations. For a two-flavor FFC computation with both spatial and temporal evolution, one can neglect the matter term due to background electron density under certain circumstances~\cite{Dasgupta:2018ulw} (see~\cite{Abbar:2017pkh} for a detailed analysis on the role of background matter), whereas in a three-flavor setup with both electrons and muons in the background, this is not trivial \cite{EstebanPretel:2007yq,Mirizzi:2009td}. However, our numerical examples only deal with time evolution and the matter terms can be safely neglected~\cite{Dasgupta:2016dbv}. Thus, only the non-linear term governs the dynamics of our setup. Note that there can be radiative corrections in this term as well, due to the presence of three-flavors, however, these are small and can be dropped~\cite{Mirizzi:2009td}.
What enters the EoMs is the difference in the occupation numbers of the different flavors integrated over energy, dubbed in literature as the corresponding neutrino flavor lepton number. In the two-flavor scenario, assuming the non-electron flavor angular distributions are identical for both the particles and the antiparticles, one defines the electron lepton number (ELN) as~\cite{Izaguirre:2016gsx}
\begin{equation}
G^e_{\bf v} = \sqrt{2} G_F \int_{0}^{\infty}\frac{dE\,E^2}{2 \pi^2}\left[\rho_{ee}(E,{\bf v})-\bar{\rho}_{ee}(E,{\bf v})\right] \,\,.
\label{eq:eln}
\end{equation}
However, 2-D simulations~\cite{Bollig:2017lki} show that this picture is not entirely accurate; the inclusion of muons does create an appreciable flux hierarchy between the $\nu_\mu$ and $\bar{\nu}_\mu$ in the accretion phase. This is primarily because muons can be pair-produced from electrons, and can participate in $\beta$-processes to create $\nu_\mu$ and $\bar{\nu}_\mu$. The relative neutron-to-proton ratio governs the extent of  muonization $(\nu_\mu-\bar{\nu}_\mu)$ in the PNS. In contrast, due to the high value of $m_\tau$, the $\tau$ density remains small throughout, although a tiny asymmetry may arise due to different scattering cross-sections with nucleons. 

Thus, within the three-flavor formalism \cite{Airen:2018nvp,Chakraborty:2019wxe}, one needs to define the corresponding muon lepton number $(\mu\,{\rm LN})$ and tau lepton number $(\tau\,{\rm LN})$ as
\begin{eqnarray}
G^\mu_{\bf v} &=&\sqrt{2} G_F \int_{0}^{\infty}\frac{dE\,E^2}{2 \pi^2}\left[\rho_{\mu\mu}(E,{\bf v})-\bar{\rho}_{\mu\mu}(E,{\bf v})\right] \,\,,\\
G^\tau_{\bf v} &=&\sqrt{2} G_F \int_{0}^{\infty}\frac{dE\,E^2}{2 \pi^2}\left[\rho_{\tau\tau}(E,{\bf v})-\bar{\rho}_{\tau\tau}(E,{\bf v})\right] \,\,.
\label{eq:mutau-ln}
\end{eqnarray}
The EoM is sensitive to the difference of these lepton numbers via the following quantities, 
\begin{eqnarray}
 G^{e\mu}_{\bf v}&=&G^e_{\bf v}-G^\mu_{\bf v} \,\,,\\
 G^{e\tau}_{\bf v}&=&G^e_{\bf v}-G^\tau_{\bf v}\,,\\
 G^{\mu\tau}_{\bf v}&=&G^\mu_{\bf v}-G^\tau_{\bf v}\,.
 \label{eq:Gv}
 \end{eqnarray}
Note that $G^{e\mu}_{\bf v}$ reduces to the ELN in the two-flavor scenario, where $n_{\nu_\mu}=n_{\bar{\nu}_\mu}$. There exists three off-diagonal elements of the occupation number matrix: $\rho_{e\mu},\,\rho_{e\tau},\,\rho_{\mu\tau}$, which might undergo an exponential growth (i.e. an instability), when a crossing is created in one of the angular distributions. Therefore, a crossing in the ELN is not enough, since one needs to consider $G^{e\mu},\, G^{e\tau}$ or $G^{\mu\tau}$ \cite{Chakraborty:2019wxe}. This has important consequences for FFC. For instance, a tiny crossing in the ELN can be erased by a negative $\mu\,{\rm LN}$. Analogously, the absence of a crossing in the ELN can be compensated by a positive $\mu\,{\rm LN}$. Similar arguments hold for the other sectors. As a result, claiming the presence of fast oscillations, focusing \emph{only} on tiny ELN crossings, as done for example in \cite{Morinaga:2020nmf}, might lead to incomplete conclusions. In what follows, we consider simple toy models to demonstrate this important point. 

\emph{Numerical examples\,}-- We assume a spatially homogeneous flavor composition, and only consider the time evolution in Eq.~\ref{eq:eom}. We assume that the initial neutrino angular distributions are axially symmetric around the $z-$axis, i.e., they only depend on the zenith angle. Nevertheless, we do consider the azimuthal angles in our calculations, so that also axially breaking instabilities are allowed to develop \cite{Raffelt:2013rqa}. We take $\mu=\sqrt{2}G_Fn_\nu=4\times10^5$ km$^{-1}$, which is a typical value in the neutrino decoupling region. Concerning the vacuum term, we use as oscillation frequencies $\Delta m^2_{31}/(2E)=0.5$ km$^{-1}$ and $\Delta m^2_{21}/(2E)=0.01$ km$^{-1}$, whereas we set the mixing angles to be $\theta_{12}=\theta_{13}=\theta_{23}=10^{-3}$. In this way, we mimic the suppression of $\theta_{ij}$ induced by the large potentials. Moreover, as we
focus only on the time evolution we neglect the matter terms, $\lambda_e =\lambda_{\mu} =\lambda_{\tau} =0$.

In the following, we consider four toy cases of $G^{e\mu}_{\bf v}$, $G^{e\tau}_{\bf v}$ and $G^{\mu\tau}_{\bf v}$,  highlighting the fact that the effective two-flavor formalism leads to different conclusions. Note that current state of the art SN simulations \cite{Bollig:2017lki} can only provide angular moments of the neutrino distribution, from which one can construct realistic angle-dependence of the neutrino flavor intensity but only near the neutrinosphere. However, these distributions are not very reliable at larger radii~\cite{privcomm}. We leave to future work the assessment of which lepton number angular distributions are realized in nature.

\begin{figure}[!t]
\begin{centering}
\includegraphics[width=0.5\textwidth]{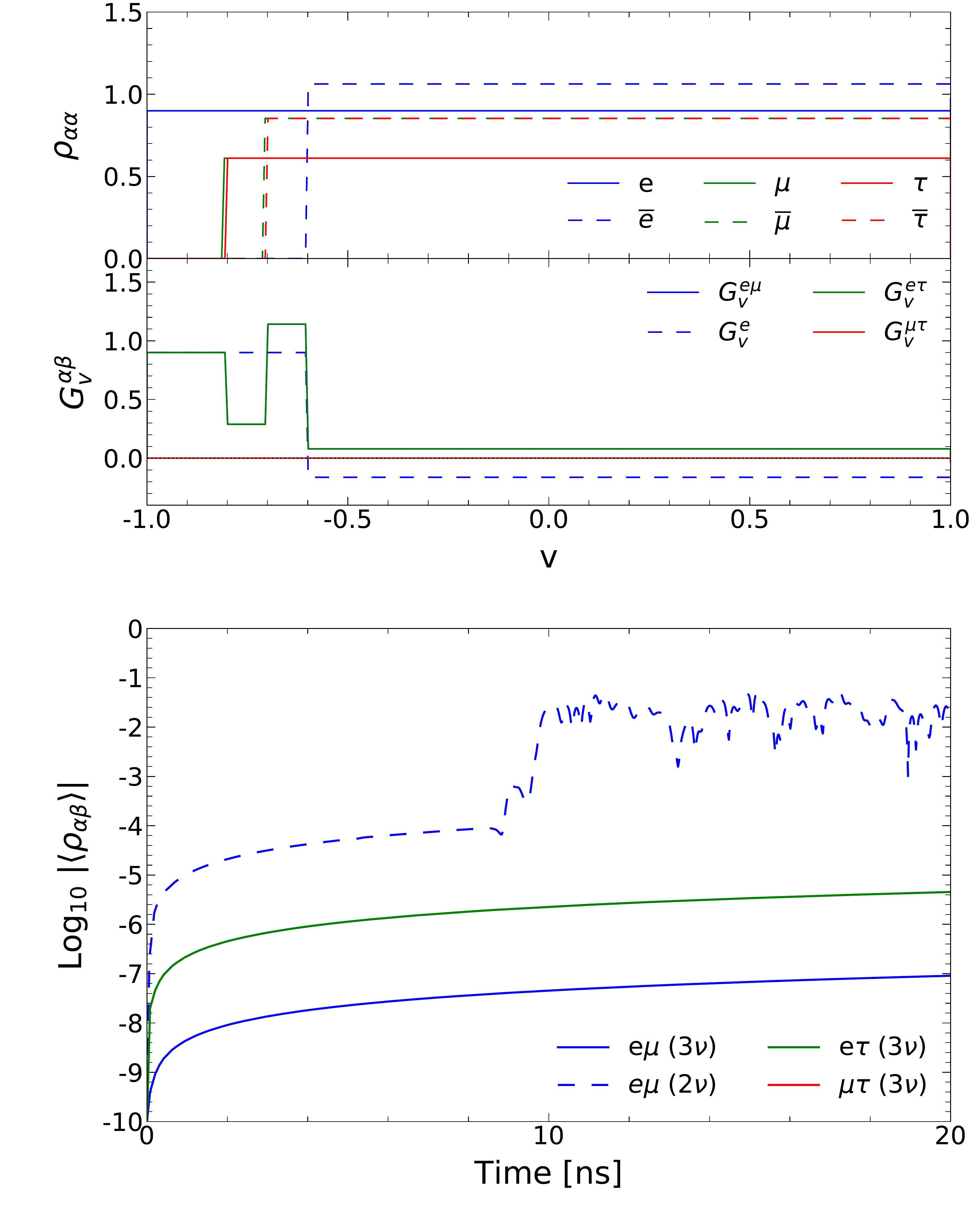} 
\end{centering}
\caption{The upper panels show the angular distribution of the fluxes and the effective lepton numbers for different flavors. The effective lepton number panel ($G^{\alpha\beta}_{\bf v}$ vs $\bf v$) shows both the three-flavor (solid lines) and the two-flavor ELN (dashed line). The lower panel shows the evolution of the angle-averaged off-diagonal elements $\left|\langle\rho_{\alpha\beta}\rangle\right|$ of the occupation number matrix with time (in ns). The three-flavor evolution (solid lines) has no instability, while the effective two-flavor evolution, assuming $\nu_x=\bar\nu_x$, shows large exponential growth (dashed line).} 
\label{fig:c1}
\end{figure}

Firstly, we consider the scenario in Fig.\ref{fig:c1}, where each of $G^{e}_{\bf v}$, $G^{\tau}_{\bf v}$ and $G^{\mu}_{\bf v}$ have crossings as
can be interpreted from the angular distribution of the fluxes (uppermost panel). We show the $G^{e}_{\bf v}$ crossing by the dashed lines in the $`G^{\alpha\beta}_{\bf v}$ vs $\bf v$' panel. Thus, in the two-flavor scenario ($n_{\nu_\mu}=n_{\bar{\nu}_\mu}$), there is an exponential growth in $\rho_{e\mu}$ (dashed line, lower panel), due to this crossing in $G^{e}_{\bf v}$.
However, it is evident that no crossing persists in $G^{e\mu}_{\bf v},\,G^{e\tau}_{\bf v}$ or $G^{\mu\tau}_{\bf v}$. Consequently, the time evolution of the off-diagonal elements of $|\langle\rho_{\alpha\beta}\rangle|$ (solid lines, lower panel) does not show any exponential growth. Here, we have defined
\begin{equation}
\left|\langle\rho_{\alpha\beta}\rangle\right|=\left|\int d{\bf v}\rho_{\alpha\beta}({\bf v})\right|\,.
\label{eq:off_diag_plot}
\end{equation} 
This simple but crucial example clearly demonstrates that some of the crossings found in \cite{Glas:2019ijo,Abbar:2019zoq,Morinaga:2020nmf} might disappear once the corresponding hydrodynamical simulation include the full three-flavor neutrino transport, including the production of muons. 

\begin{figure}[!t]
\begin{centering}
\includegraphics[width=0.5\textwidth]{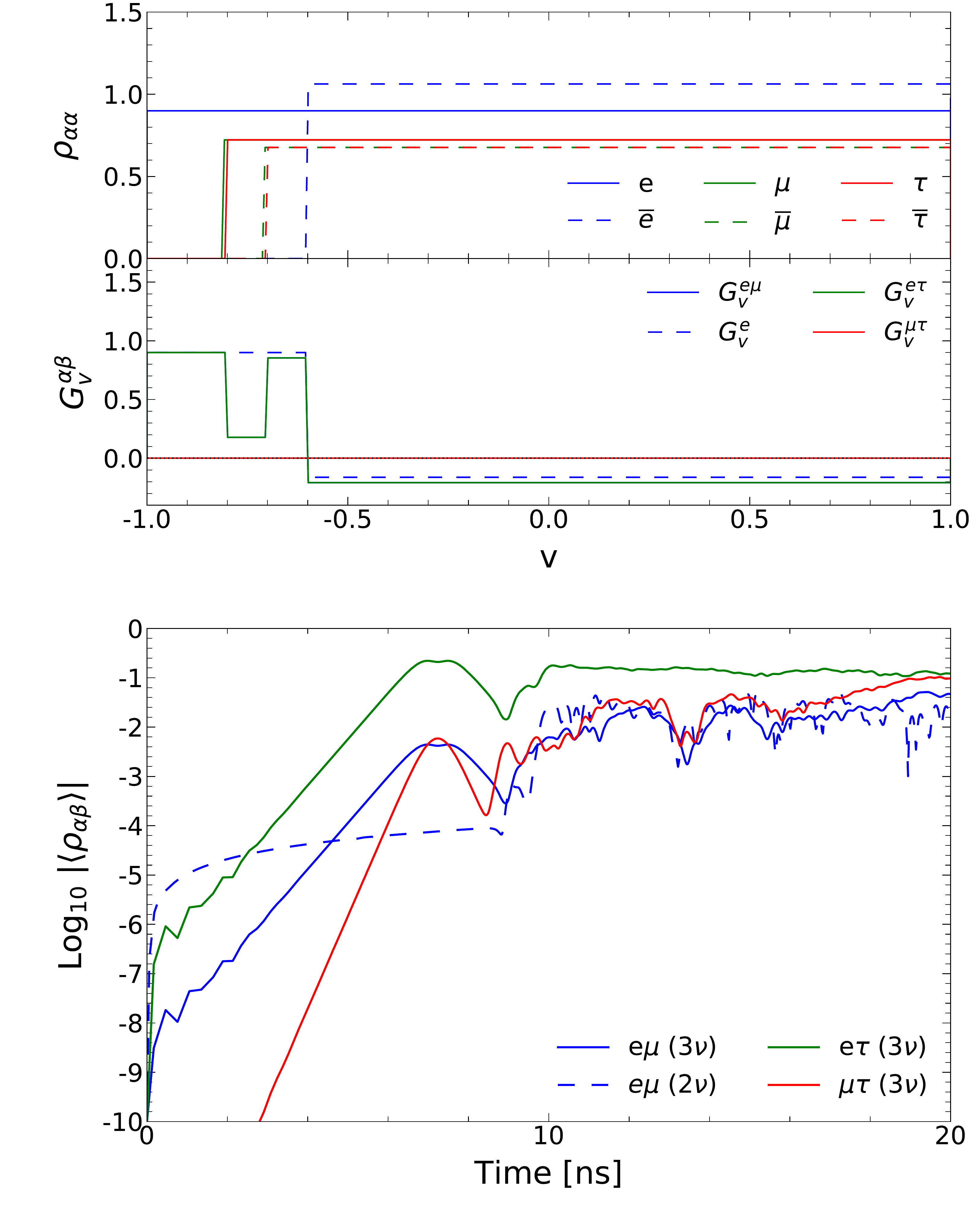} 
\end{centering}
\caption{Panels represent same as in Fig.\,\ref{fig:c1}. However, due to large crossings in $G^{e\mu}_{\bf v},\,G^{e\tau}_{\bf v}$, substantial flavor conversions are seen in all the sectors.} 
\label{fig:c2}
\end{figure}
\vspace{0.2cm}  
As a second example, consider Fig.\,\ref{fig:c2} (upper panels), where the ELN ($G^{e}_{\bf v}$) has a regular crossing (dashed line), but there are none in the $\mu\,{\rm LN}$ or the $\tau\,{\rm LN}$. The spectra are designed such that the flavor lepton number difference $G^{e\mu}_{\bf v},\,G^{e\tau}_{\bf v}$ exhibit  deep crossings (solid lines). On the other hand, the crossing in $G^{\mu\tau}_{\bf v}$ is extremely shallow. The naive two-flavor intuition is that there should be exponential growths in the $e-\mu$ and $e-\tau$ sector. However, the non-linear, coupled nature of the problem intertwines the growths in all the sectors. This is borne by the lower panel, where substantial flavor conversion is seen in all the sectors, though with different growth rates. 
Note that the growth in $\rho_{\mu\tau}$ is inherently a non-linear effect, and will not be captured by a linear stability analysis.
The corresponding two-flavor evolution of $\rho_{e\mu}$ is also shown for comparison. Clearly, the two-flavor evolution is very different, with a larger onset time and growth rate.
\begin{figure}[!t]
\begin{centering}
\includegraphics[width=0.5\textwidth]{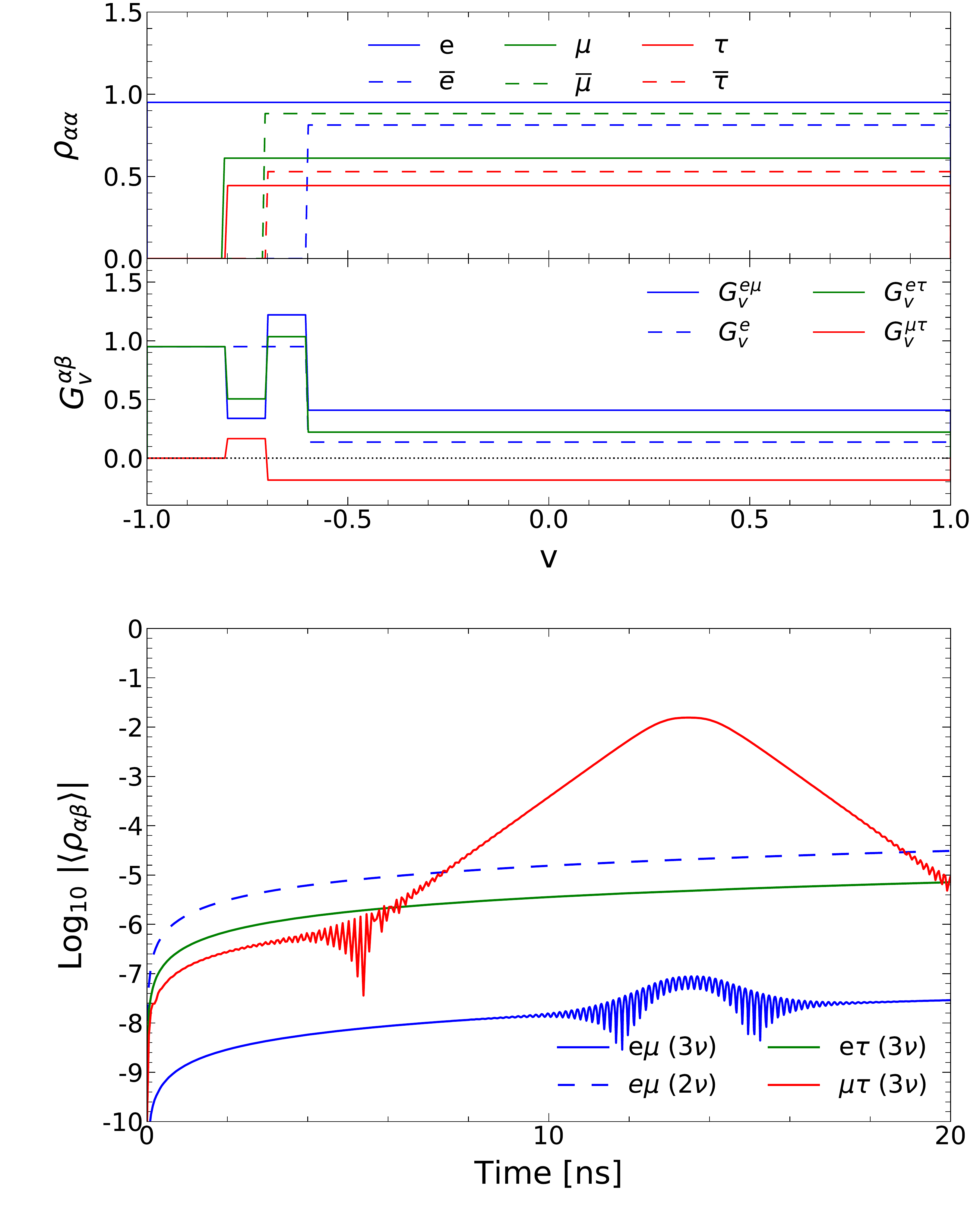}
\end{centering}
\caption{Panels represent same as in Fig.\,\ref{fig:c1}. Crossing only in $G^{\mu\tau}_{\bf v}$ causes an exponential growth in $\mu-\tau$ sector (lower panel). The two-flavor evolution shows no instability.} 
\label{fig:c3}
\end{figure}

The third case, shown in Fig.\,\ref{fig:c3}, presents a crossing only in $G^{\mu\tau}_{\bf v}$. There exists a reasonable asymmetry (upper panels) between $\nu_\mu$ and $\nu_\tau$ (and between their antiparticles as well) to generate an exponential growth in the $\mu-\tau$ sector. This is what is seen in the lower panel, where the  $\mu-\tau$ sector experiences a flavor instability, while the other two do not. Indeed, the two-flavor analysis (dashed line) also does not exhibit any instability due to the lack of a crossing in the ELN ($G^{e}_{\bf v}$).
This example advances the hypothesis that those regions where no ELN crossing was found in \cite{Glas:2019ijo,Abbar:2019zoq,Morinaga:2020nmf} might, in reality, have fast instabilities once the differences between $\nu_\mu$ and $\bar{\nu}_\mu$ are taken into account. Another comment is in order: the amplitude of the exponential growth in our toy model is not enough to cause substantial flavor conversions. However, the background conditions for these solutions may dynamically change in a realistic SN environment, or if spatial evolution is taken into account, and may result in flavor conversions.

\begin{figure}[!t]
\begin{centering}
\includegraphics[width=0.5\textwidth]{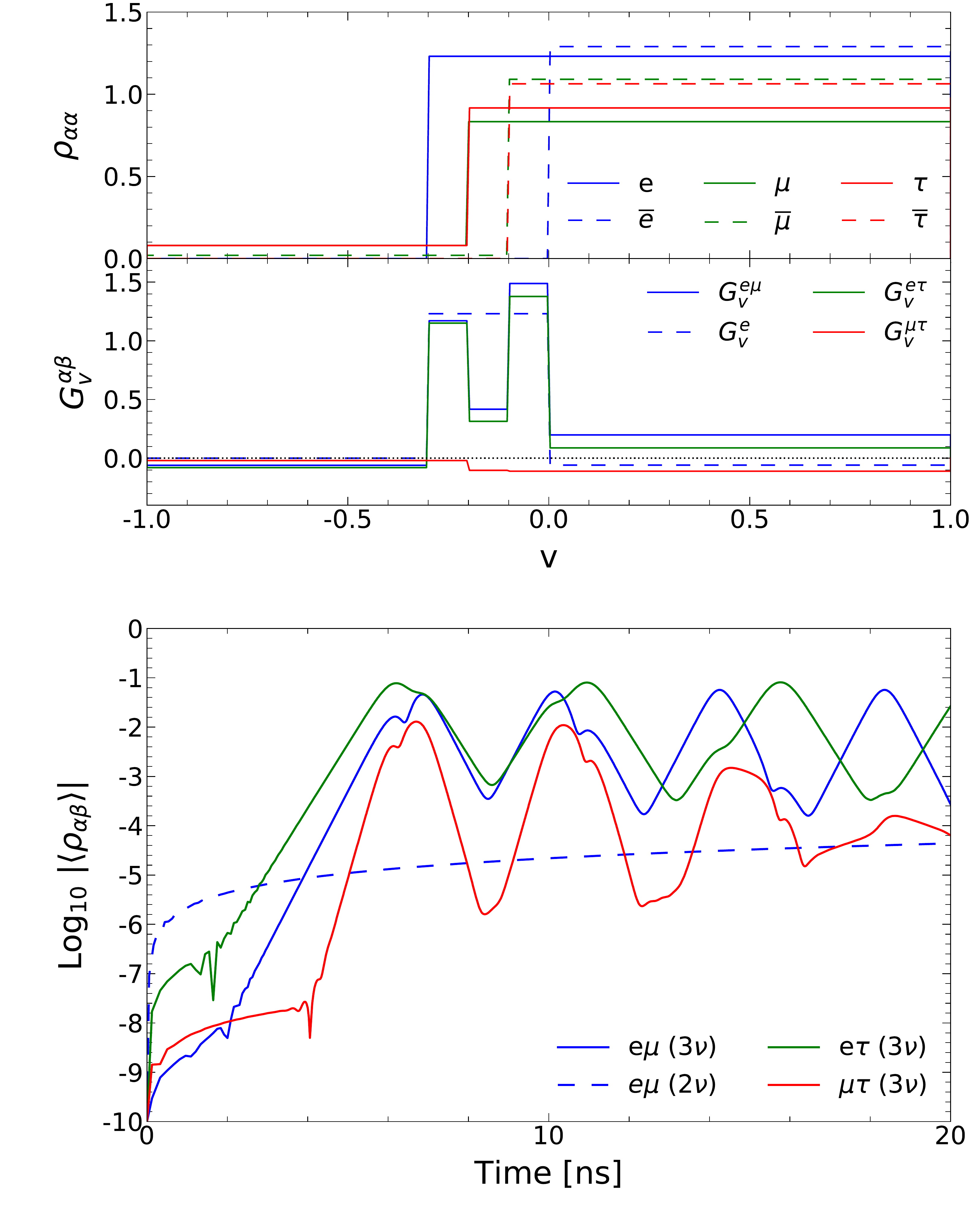} 
\end{centering}
\caption{Panels represent same as in Fig.\,\ref{fig:c1}. However, shallow crossings are present in $G^{e\mu}_{\bf v},\,G^{e\tau}_{\bf v}$ in the backward direction (upper panels), leading to exponential growths in  $\left|\langle\rho_{\alpha\beta}\rangle\right|$ (lower panel) for the three-flavor setup. The two-flavor evolution shows no instability.} 
\label{fig:c4}
\end{figure}

As a final example, we consider a scenario in Fig.\,\ref{fig:c4} where shallow crossings are present in $G^{e\mu}_{\bf v},\,G^{e\tau}_{\bf v}$ in the backward direction, whereas $G^{\mu\tau}_{\bf v}$ shows a significant crossing in the forward direction as well. To contrast with the two-flavor examples, the setup is constructed such that the ELN ($G^{e}_{\bf v}$) also has a shallow crossing but in the forward direction only. We find that such shallow crossings readily lead to an instability in the three-flavor case, whereas the two-flavor setup shows no instability (lower panel). This example is motivated from~\cite{Morinaga:2020nmf}, where it was pointed out that shallow crossings in the backward directions can lead to a fast instability.
In \cite{Morinaga:2020nmf}, such backward crossings were associated with residual coherent scattering on heavy nuclei, which is slightly enhanced for $\bar{\nu}_e$ with respect to $\nu_e$, because of their larger average energy. Our toy model advances the hypothesis that the existing differences between $\nu_\mu$ and $\bar{\nu}_\mu$ could (at least in principle) be the real cause of these crossings. Similarly, the non negligible muon lepton number can also erase a potential shallow (backward) crossing in the ELN. Our examples clearly establish the importance of detailed three-flavor treatments to assess whether such possibilities are indeed realized in nature. We leave this task to future work.

\emph{Discussion and conclusions.\,}--
Fast neutrino flavor conversion near the SN core can lead to a paradigm change in our understanding of flavor evolution of supernova neutrinos. Within a two-flavor formalism, these ultra-rapid flavor conversions are believed to occur mainly when the ELN, i.e., the difference between the $\nu_e$ and $\bar{\nu}_e$ angular spectra, exhibit a zero-crossing. Such flavor mixing can have a drastic impact on the shockwave revival, as well as nucleosynthesis. Hence, it is crucial to appreciate these flavor conversions, using a complete three-flavor analysis.

We have performed, for the first time, a completely non-linear, three-flavor treatment of fast flavor conversions of neutrinos. We find that the inclusion of three-flavors can significantly alter our understanding of the conditions for fast conversions. Using simple toy spectra, we demonstrate that it is not the ELN, or correspondingly, the $\mu$LN, and $\tau$LN, but rather their differences that govern these fast modes. The examples studied in this paper clearly show that three-flavor evolution can result in instabilities that are not captured by a two-flavor study; conversely, it can also wash out the instabilities predicted by a two-flavor study. These results also show that the linear stability analysis cannot capture all the instability signatures of the full non-linear analysis, as expected.

Our findings further indicate caution against claiming the presence of fast conversions for shallow crossings in the ELN, because such crossings can easily be nullified by an opposite crossing in the $\mu$LN. 
This is particularly relevant in the shock region, where a significant population of $\nu_\mu,\nu_\tau$ can be expected in the accretion phase. This motivates the necessity of including muons in a dedicated analysis of fast-flavor conversions to gauge their impact on supernova dynamics.

\emph{Acknowledgments\,}--We would like to thank Georg Raffelt, Alessandro Mirizzi, Basudeb Dasgupta and Sajad Abbar for useful comments on the manuscript. MS acknowledges support from the National Science Foundation, Grant PHY-1630782, and to the Heising-Simons Foundation, Grant 2017-228. The work of F.C. is supported by the Deutsche Forschungsgemeinschaft through Grants SFB-1258 “Neutrinos and Dark Matter in Astro- and Particle Physics (NDM)” and EXC 2094 “ORIGINS: From the Origin of the Universe to the First Building Blocks of Life”. SC acknowledges the support of the Max Planck India Mobility Grant from the Max Planck Society. MC and SC also recived funding/support from the European Union’s Horizon 2020 research and innovation programme through the InvisiblesPlus RISE under the Marie Skłodowska-Curie grant agreement No 690575 and through the Elusives ITN  under the Marie Skłodowska -Curie grant agreement No 674896.

\bibliographystyle{JHEP}
\bibliography{biblio}

\end{document}